\documentstyle[twoside,fleqn,espcrc2]{article}

\newcommand{\AmS}{{\protect\the\textfont2
  A\kern-.1667em\lower.5ex\hbox{M}\kern-.125emS}}

\title{Blockspin Scheme and Cluster Algorithm for
       Quantum Spin Systems}

\author{U.-J. Wiese\address{Institut f\"ur Theoretische Physik, Universit\"at
        Bern, Sidlerstrasse 5, CH-3012 Bern, Switzerland }
        and H.-P. Ying$^{\rm a}$
           \address{Zhejiang
        Institute of Modern Physics, Zhejiang University,
        Hangzhou 310027, P.R.China }\thanks{speaker at the conference}}
\begin{document}

\begin{abstract}
We present a numerical study using a cluster algorithm for the 1-d $S=1/2$
quantum Heisenberg models. The dynamical critical
exponent for anti-ferromagnetic chains is $z=0.0(1)$ such that
critical slowing down is eliminated.
\end{abstract}

\maketitle

\section{INTRODUCTION}

Since 1976 it is well-known that several mappings of the partition function
of a $d$-dimensional $S=1/2$ quantum spin model onto a $(d+1)$-dimensional
classical spin model exist \cite{Suzu76,Cull83}.
The maps lead to a Feyman path integral formula of quantum
statistical mechanics. The equivalence opens the possibility to use
powerful computational techniques to obtain information on quantum spin
systems. Monte Carlo (MC) simulation, a method to get precise numerical data,
has played an important role in the studies \cite{Raed85}, and has especially
been applied to the anti-ferromagnetic models \cite{Uchi85,Barn91} since
Haldane's conjecture \cite{Hald83} and the discovery of high-$T_{c}$
superconductors \cite{Bedn86}. There are, however, some problems with the
standard numerical methods, which use importance-sampling techniques for
the $(d+1)$-dimensional induced classical spin
systems which have multi-spin interactions. Their effect is characterized
by large autocorrelation times (in units of MC sweeps needed to
create a new statistically independent configuration) in the low temperature
region and in the euclidean time continuum limit. In this paper a cluster
algorithm is developed for simulations of quantum spin systems\cite{Wies92},
especially of 1-d quantum Heisenberg models, after mapping them
to 2-d induced classical spin models and using a blockspin
scheme.

\section{METHOD}

Consider for definiteness a 1-d $S=1/2$ Heisenberg model defined on a chain
with an even number of spins $L=2N$ (fig.1) with periodic boundary conditions.
The Hamilton operator is given by
\begin{equation}
     H= \frac{J}{4} \sum_{i=1}^L {\vec\sigma}_{i} \cdot {\vec\sigma}_{i+1}.
\end{equation}
\begin{figure}
\vspace{23mm}
\caption{Heisenberg magnet on a chain with an even number of spins $2N$.}
\end{figure}
Here $\vec{\sigma}_{i}$ is a Pauli spin operator at the point $i$ on the
chain. $J>0$ corresponds to an anti-ferromagnet (AF), while $J<0$ corresponds
to a ferromagnet (F). Using the checkerboard decomposition
$H = H_1 + H_2$ (fig.1) and the Trotter formula \cite{Raed85},
the partion function is expressed as a path integral given by

\begin{equation}
     Z=\mbox{Tr}\exp(-\beta H)=\prod_{x,t}\sum_{s(x,t)=
     \pm 1}\exp(-S),
\end{equation}
where $\beta$ is the inverse temperature. Eq.(2) is equivalent to a 2-d
classical spin model on a rectangular lattice $L\times 2M$ with Ising-like
variables $s(x,t)= \pm 1$. The lattice spacing in the additional
so-called euclidean time or Trotter direction is
$\epsilon=1/M$ (fig.2). The classical spins interact with each other via
4-spin couplings $\exp(-S[s_{1},s_{2},s_{3},s_{4}])=
\langle s_{1},s_{2}|\exp(-\epsilon\beta J
{\vec\sigma}_{i} \cdot {\vec\sigma}_{i+1}/4)|s_{3},s_{4}\rangle$, which are
given by elements of the transfer matrix. For the AF one has
\begin{eqnarray}
&&S[1,1,1,1]=S[-1,-1,-1,-1]=\epsilon\beta J/4, \nonumber \\
&&S[1,-1,1,-1]=S[-1,1,-1,1]= \nonumber \\
&&\qquad \epsilon\beta J/4-\log [(\exp(\epsilon\beta  J)+1)/2], \nonumber \\
&&S[1,-1,-1,1]=S[-1,1,1,-1] = \nonumber \\
&&\qquad \epsilon\beta J/4-\log [(\exp(\epsilon\beta J)-1)/2].
\end{eqnarray}
All other action values are infinite and the corresponding elements
of the transfer matrix are zero. The matrix elements can be
interpreted as the Boltzmann weights of the spin configurations.
Up to here the quantum spins ${\vec\sigma}_i$ with 2-spin couplings living
on a 1-d chain have been mapped to classical spins $s(x,t)$ with 4-spin
couplings living in the 2-d plane. However, most classical spin
configurations are
forbidden, because for them some elements of the transfer matrix are zero
(their Boltzmann factor vanishes). Therefore, it is natural to attempt to
choose a collective nonlocal update technique. For models with 2-spin
couplings this can be done using the Swendsen-Wang \cite{Swen87} or Wolff
\cite{Wolf89} cluster algorithms
which flip whole cluster of spins simultaneously. These algorithms, however,
can not be applied directly to a model with 4-spin couplings.\\
\begin{figure}[htp]
\vspace{66mm}
\caption{2-d induced classical spin system with 4-spin couplings
         depicted by shaded plaquettes. (a) The scheme $\{B\}$:
         each blockspin consists of 4 spins and interacts with its
         nearest-neighbors via the original 4-spin
         interaction. (b) the other scheme $\{\bar{B}\}$ which is
         dual to the scheme $\{B\}$.}
\end{figure}
  To make an application of the cluster technique  possible we further map the
classical spin model with 4-spin couplings $S[s_{1},s_{2},s_{3},s_{4}]$
to a blockspin model. A blockspin $B(n,m)=
\{ s(x,t),s(x+1,t),s(x,t+1),s(x+1,t+1)\}$ is a collection of four spins,
 where $x=2n-1,t=2m$. The blockspins interact via
2-blockspin
interactions $S[B,B']=S[s_1,s_2,s_3,s_4]$ as shown in fig.2a. Each spin
belongs exactly to one blockspin and the blockspins live on a lattice with a
doubled lattice spacing. The spins can also be arranged to blockspins in
another way, $\bar{B}(n,m)=\{ s(x,t),s(x+1,t),s(x,t+1),s(x+1,t+1)\}$ with
$x=2n,t=2m-1$ (see fig.2b). An updating algorithm for the blockspins must
alternate between the schemes $\{B\}$ and $\{\bar{B}\}$ to ensure
ergodicity. We have also chosen two kinds of nonlocal blockspins \cite{Wies92}
to change the magnetization and the so-called winding number
by odd integers.
The cluster algorithm makes use of a flip symmetry of the blockspin model. A
blockspin $B=\{ s_{1},s_{2},s_{3},s_{4}\}$ is flipped to $-B=\{ -s_{1},-s_{2},
-s_{3},-s_{4} \}$ simply by flipping all spins in $B$. The
blockspin action $S[B,B']$ is invariant under a flip of all blockspins
because the original action is symmetric against flipping all spins.
\begin{figure}[htb]
\vspace{119mm}
\caption{A $\log$-$\log$ plot of the autocorrelation time $\tau_{e}$ versus
         $1/\epsilon$ at $\beta J=1$ both for $(a)$ AF
         and $(b)$ F  Heisenberg magnets with $2N=32$.}
\end{figure}
\section{MONTE CARLO RESULTS}
  We now describe how to generate transitions between blockspin
configurations.
$a$) The algorithm puts bonds between all nearest-neighbor pairs of blockspins
      with a probability
     $p=1-\mbox{min} \{1,\exp(-S[-B,B'])/\exp(-S[B,B'])\}$.
$b$) Two blockspins belong to the same cluster if they are connected by a bond.
$c$) All blockspins in one cluster must be flipped simultaneously.
$d$) Using the Swendsen-Wang multi-cluster method \cite{Swen87} each cluster
     is flipped
     independently with probability 1/2. For the Wolff single-cluster
     algorithm \cite{Wolf89} one blockspin is randomly chosen and the cluster
to  which it belongs is flipped.
The procedure $a$)-$d$) of changing a configuration satisfies the
detailed balance condition \cite{Wolf89}. For ergodicity we alternate between
the  $\{B\}$ and $\{\bar{B}\}$ schemes.
We tested the blockspin cluster algorithm in detail for one-dimensional spin
chains. To demonstrate the efficiency of the algorithm we
compare it to a blockspin Metropolis update. For both algorithms
we measure the autocorrelation functions $C_{\cal O}(t)$, where $\cal O$
denotes the observables, i.e. the internal energy $e$, and the uniform and
staggered susceptibilities $\chi$ and $\chi_{s}$. Then we obtain the
integrated autocorrelation times $\tau_{\cal O}$ by
\begin{equation}
\exp(-1/\tau_{\cal O})=\sum_{t=1}^{\infty}C_{\cal O}(t)/
\sum_{t=0}^{\infty}C_{\cal O}(t).
\end{equation}
   In all the simulations we have performed a random start followed
by 5000 sweeps for thermalization and 50000 sweeps for measurements using
the single-cluster updating.
   From the data at $\beta J=1$ shown in fig.3 we obtain the dynamical
critical exponent $z_{e}$ both for the anti-ferromagnet and for the
ferromagnet. $z_{e}$ is defined in the continuum limit $\epsilon \rightarrow 0$
by $\tau_{e} \propto 1/\epsilon^{z}$. We find
\begin{eqnarray}
z_{e}&=&0.0(1) \mbox{\quad for the cluster algorithm},\nonumber \\
z_{e}&=&0.8(1) \mbox{\quad for the Metropolis algorithm}.
\end{eqnarray}
The autocorrelation times  $\tau_{\chi}$ and $\tau_{\chi_{s}}$
do not diverge in the continuum limit even for the Metropolis algorithm.
Still, the corresponding autocorrelation times are at least an order of
magnitude larger than the ones of the cluster algorithm.\\
As shown in fig.4, the Metropolis algorithm has severe problems with
slowing down at low temperatures. At $\beta J=8$ for example
$\tau_{\chi_s}= 12(1)$ for $J>0$, and
$\tau_{\chi}=3300(200)$ for $J<0$.
The cluster algorithm, on the other hand, has autocorrelation times of at
most a few sweeps and it can update the configurations efficiently at
 lower temperatures like e.g. $\beta J=16$.
 In fig.4 the $\tau_{\chi}$ and $\tau_{\chi_s}$ are fitted by
\begin{equation}
\tau_{\chi} \propto \exp(c_{\chi}\beta |J|), \quad
\tau_{\chi_s} \propto \exp(c_{\chi_s}\beta |J|).
\end{equation}
The data at $\beta J=16$ show some
slowing down for the ferromagnet, but the autocorrelation times are moderate.
For anti-ferromagnet there is no indication of slowing down such that
\begin{equation}
z_{e} = z_{\chi} =z_{\chi_{s}} = 0.0(1).
\end{equation}
\begin{figure}[htb]
\vspace{119mm}
\caption{$(a)$ The plot of $\log \tau_{\chi_{s}}$ versus $\beta J$ for
         AF couplings and $(b)$ $\log \tau_{\chi}$ for F couplings with
         $2N=128$ and $\epsilon=\beta /M=0.125$.}
\end{figure}
\section{DISCUSSION}
First, we like to  mention a few words about frustration. As it is well known
the Swendsen-Wang and Wolff cluster algorithms eliminate critical slowing
down for the Ising and Potts models because they are not frustrated. For
strongly frustrated models, e.g. for spin glasses, cluster algorithms do not
work efficiently. Therefore the question arises if our blockspin models are
frustrated or not. We have verified that the blockspin model for 1-d AF
couplings has indeed no frustrated allowed configurations. For the 1-d
ferromagnet, on the other hand, some allowed configurations have a weak
frustration. However, in the continuum limit $\epsilon\rightarrow 0$
the frustration disappears. We believe that the weak frustration is
responsible for the mild slowing down in the ferromagnetic case at low
temperature.

   The blockspin cluster algorithm can also be applied to other models, e.g.
to systems with anisotropic couplings or in an external magnetic field, to
models of higher spins ($s=1,3/2,...$), to six- and eight-vertex models
and to 1-d fermion
systems. Work on the 2-d anti-ferromagnet is in progress \cite{Wies93}.

  We are grateful to P. Hasenfratz for useful suggestions and discussions.
This work was supported partly by the Schweizer Nationalfond. One of us (HPY)
would like to acknowledge the support from the PAO's scholarship for CSSA.


\begin{thebibliography}{9}
\bibitem{Suzu76} M. Suzuki, Prog. Theor. Phys. 56 (1976) 1454;
                 Comm. Math. Phys. 51 (1976) 183.
\bibitem{Cull83} J. J. Cullen and D. P. Landau, Phys. Rev. B27 (1983) 297.
\bibitem{Raed85} H. De. Raedt and A. Lagendijk, Phys. Rep. 127 (1985) 233.
\bibitem{Uchi85} M. Uchinami, Phys. Rev. B39 (1989) 4554; K. Normura,
                 Phys. Rev. B40 (1989) 2421.
\bibitem{Barn91} T. Barnes, Int. J. Mod. Phys. C2(1991)659;
                 M. S. Makivic and H.-Q. Ding, Phys.Rev. B43 (1991) 3562.
\bibitem{Hald83} F. D. M. Haldane, Phys. Rev. Lett. 50 (83) 1153;
                 Phys. Lett. 93A (1983) 464.
\bibitem{Bedn86} G. Bednorz and K. A. M\"uller, Z. Phys. B64 (1986) 189.
\bibitem{Wies92} U.-J. Wiese and H.-P. Ying, Phys. Lett. 168A (1992) 143.
\bibitem{Swen87} R. Swendsen and J.-S. Wang, Phys. Rev. Lett. 58 (1987) 86.
\bibitem{Wolf89} U.Wolff, Phys. Rev. Lett. 62 (1989) 361;
                 Nucl. Phys. B334 (1990) 581.
\bibitem{Wies93} U.-J. Wiese and H.-P. Ying, in preparation.
\end{thebibliography}
\end{document}